\documentclass[10pt, pre, eps, twocolumn,showpacs,nofootinbib]{revtex4-1}
\usepackage{url,ulem}
\usepackage{fourier}
\usepackage{amssymb,amsmath}
\usepackage[dvipdfmx]{graphicx}
\usepackage{graphicx,color}

\newcommand{\dfracp}[2]{\dfrac{\partial #1}{\partial #2}}
\newcommand{\ave}[1]{\left\langle #1 \right\rangle}
\newcommand{\norm}[1]{\left|\left| #1 \right|\right|}
\newcommand{\Heqz}{H_{0}}
\newcommand{\Veqz}{V_{0}}
\newcommand{\feqz}{f_{0}}

\newcommand{\bc}{\boldsymbol{c}}
\newcommand{\bh}{\boldsymbol{h}}
\newcommand{\bbm}{\boldsymbol{m}}

\begin{document}
\title{Nondivergent and negative susceptibilities around critical points
  of a long-range Hamiltonian system with two order parameters}
\author{Yoshiyuki Y. Yamaguchi}
\email{yyama@amp.i.kyoto-u.ac.jp}
\author{Daiki Sawai}
\affiliation{
Department of Applied Mathematics and Physics, 
Graduate School of Informatics, Kyoto University, 
606-8501, Kyoto, Japan}
%\today
\begin{abstract}
  The linear response is investigated in a long-range Hamiltonian system
  from the view point of dynamics, which is described by the Vlasov equation
  in the limit of large population.
  Due to existence of the Casimir invariants of the Vlasov dynamics,
  an external field does not drive the system to the forced thermal equilibrium
  in general, and the linear response is suppressed.
  With the aid of a linear response theory based on the Vlasov dynamics,
  we compute the suppressed linear response 
  in a system having two order parameters,
  which introduce the conjugate two external fields
  and the susceptibility matrix of size two accordingly.
  Moreover, the two order parameters bring three phases
  and the three types of second-order phase transitions between two of them.
  For each type of the phase transitions,
  all the critical exponents for elements of the susceptibility matrix
  are computed.
  The critical exponents reveal that some elements of the matrices
  do not diverge even at critical points,
  while the mean-field theory predicts divergences.
  The linear response theory also suggests appearance of negative
  off-diagonal elements,
  in other words, an applied external field decreases
  the value of an order parameter.
  These theoretical predictions are confirmed by direct numerical
  simulations of the Vlasov equation.
\end{abstract}

\maketitle 

\section{Introduction}

The phase transition is one of the central issues
in the field of many-body systems.
It is classified into some universality classes,
and in particular, the mean-field universality class 
is easily understood by the Landau's phenomenological theory
\cite{landau-37}.
Nevertheless, a new aspect of the mean-field universality
class is recently revealed by considering dynamics.

Dynamics of the mean-field class,
including the systems having long-range interaction
\cite{campa-giansanti-moroni-00,mori-10,campa-dauxois-ruffo-09},
is described by the VLF's equation,
or the collisionless Boltzmann equation,
in the limit of large population
\cite{braun-hepp-77,dobrushin-79,neunzert-84,spohn-91}.
The Vlasov equation has the infinite number of Casimir invariants,
and these invariants may prevent the system from relaxing to
thermal equilibrium.
Indeed, when the initial state has different values of the Casimir
invariants from ones in thermal equilibrium,
then the relaxation is impossible.
We note that, with finite population, the finite-size fluctuation
plays the role of collision and drives the system to thermal equilibrium,
while the relaxation time gets longer as the population increases
\cite{zanette-montemurro-03,yamaguchi-04,barre-06,moyano-anteneodo-06,jain-bouchet-mukamel-07,debuyl-mukamel-ruffo-11,chavanis-12}.

The Casimir invariants hold even when an external field is applied,
and the invariants suppress the response \cite{mazur-69,suzuki-71}.
This suppression may induce reduction of the critical exponent
for the linear response in the Vlasov dynamics.
In a ferromagnetic body, the critical exponents $\gamma_{\pm}$
of susceptibility $\chi$ are defined as
$\chi\propto\tau^{-\gamma_{\pm}}$
around the second order phase transition.
Here $\tau$ is the parameter distance from the critical point
like $|T-T_{\rm c}|$ with temperature $T$ and its critical value $T_{\rm c}$,
and $\gamma_{+}$ ($\gamma_{-})$ is defined 
in the paramagnetic (ferromagnetic) phase.
The classical values of $\gamma_{\pm}$ in the mean-field universality
class are $\gamma_{\pm}=1$.
However, in the Vlasov dynamics of the Hamiltonian mean-field (HMF) model
\cite{inagaki-konishi-93,antoni-ruffo-95},
which is a paradigmatic toy model of a ferromagnetic body
in the mean-field class, the linear response theory
gives $\gamma_{+}=1$ \cite{patelli-gupta-nardini-ruffo-12,ogawa-yamaguchi-12}
but $\gamma_{-}=1/4$ \cite{ogawa-patelli-yamaguchi-14}.
The nonclassical critical exponent is not restricted in the HMF model,
and the universality is discussed for spatially
periodic one-dimensional systems \cite{ogawa-yamaguchi-15}.

In the HMF model, detection of nonclassical critical exponents is extended
to the nonlinear response at the critical point \cite{ogawa-yamaguchi-14}
and to the correlation length \cite{yamaguchi-16},
which is generalized to the infinite-range models
by introducing the coherent number of particles
\cite{botet-jullien-pfeuty-82,botet-jullien-83}.
Interestingly, the nonclassical critical exponents share
some scaling relations with the classical critical exponents.

Another direction of detecting nonclassical critical exponents
is to consider the linear response in extended models.
In this article, we consider the so-called
generalized Hamiltonian mean-field (GHMF) model \cite{teles-12}.
In the HMF model, particles are confined on the unit circle,
and interaction potential consists of the spatial first Fourier mode only.
Introducing the second Fourier mode, the GHMF model acquires
the Nematic phase in addition to the paramagnetic (Para)
and the ferromagnetic (Ferro) phases.
As a result, the GHMF model has the new two phase transitions:
the Para-Nematic and the Nematic-Ferro phase transitions.
As observed in the HMF model, the critical exponents in the Vlasov dynamics
may differ between the two sides of a phase transition,
and hence we need to consider six sides for the three phase transitions.
Moreover, the susceptibility in one side is described by a $2\times 2$ matrix,
since the three phases are characterized by the two order parameters
corresponding to the two Fourier modes
and each order parameter has the conjugate external field.
Consequently, we must consider $6$ critical exponent matrices
of the size $2\times 2$ and the total number of $\gamma$ is $24$ accordingly.

Appearance of the Nematic phase and the matrix form of the susceptibility
give natural questions:
Does the appearance of the Nematic phase drastically change
the critical exponents from the HMF model?
Are there any differences in the off-diagonal elements of the critical
exponent matrix between the mean-field theory and the Vlasov dynamics?
The purpose of this paper is to answer to these questions.
We compute the $24$ critical exponents theoretically
by using a response theory based on the Vlasov dynamics
\cite{ogawa-yamaguchi-15},
and confirm theoretical predictions by performing direct numerical simulations
of the Vlasov dynamics.
In the HMF model the reduction of the critical exponent is observed,
but we show a stronger result in the GHMF model
that some elements of the susceptibility matrices
do not diverge at the critical point,
even they diverge in the mean-field theory.
Further, close to the critical point,
we demonstrate that the off-diagonal elements of susceptibility matrix
become negative in the Ferro side of the Para-Ferro phase transition.

This article is constructed as follows.
The GHMF model and the three phases
are introduced in Sec.\ref{sec:GHMF}
with the corresponding Vlasov equation.
Responses in statistical mechanics and in the Vlasov dynamics
are derived in Secs.\ref{sec:statistical-mechanics}
and \ref{sec:dynamics}, respectively.
Theoretical predictions are examined numerically in Sec.\ref{sec:numerics}.
The last section \ref{sec:summary} is devoted to a summary and discussions.

\section{Generalized Hamiltonian mean-field model}
\label{sec:GHMF}

\subsection{The model}
The GHMF model represents particles confined on the unit circle
and is described by the Hamiltonian
\begin{equation}
  \label{eq:HN}
  \begin{split}
    H_{N} & = \sum_{j=1}^{N} \dfrac{p_{j}^{2}}{2}
    + \dfrac{1}{2N} \sum_{j,k=1}^{N} \Phi(q_{j}-q_{k})
    - \sum_{a=1}^{2} h_{a} \sum_{j=1}^{N} \cos aq_{j}.
  \end{split}
\end{equation}
The position of $j$-th particle is $q_{j}\in (-\pi,\pi]$,
and $p_{j}\in\mathbb{R}$ is the conjugate momentum.
$h_{1}$ and $h_{2}$ represent strength of the external fields.
The interaction potential $\Phi$ is
\begin{equation}
  \Phi(q) =  1 - \Lambda_{1} \cos q - \Lambda_{2}\cos 2q
\end{equation}
where $\Lambda_{1}$ and $\Lambda_{2}$ are non-negative constants.
Setting $\Lambda_{1}=1$ and $\Lambda_{2}=0$, and restricting $h_{2}=0$,
the GHMF model results to the HMF model with the external field $h_{1}$.
The coefficients are originally defined as $\Lambda_{1}=\Delta$ and $\Lambda_{2}=1-\Delta$
with $\Delta\in [0,1]$ to ensure the attractive interaction,
but we slightly restrict the parameter interval
as $\Lambda_{1}\in (0,1)$ and $\Lambda_{2}=1-\Lambda_{1}\in (0,1)$ for later convenience.
Corresponding to the two Fourier modes in $\Phi(q)$,
the two order parameter vectors are defined as
\begin{equation}
  \dfrac{1}{N} \sum_{j=1}^{N} (\cos q_{j}, \sin q_{j}),
  \quad
  \dfrac{1}{N} \sum_{j=1}^{N} (\cos 2q_{j}, \sin 2q_{j}),
\end{equation}
but we may set the sine parts to be zeros from the rotational
symmetry of the system and omit them accordingly.
The remaining parts,
\begin{equation}
  M_{1} = \dfrac{1}{N} \sum_{j=1}^{N} \cos q_{j}, \quad
  M_{2} = \dfrac{1}{N} \sum_{j=1}^{N} \cos 2q_{j},
\end{equation}
are conjugate to the external field $h_{1}$ and $h_{2}$ respectively.

In the limit $N\to\infty$, 
dynamics is described by the Vlasov equation
\begin{equation}
  \label{eq:Vlasov}
  \dfracp{f}{t} + \dfracp{\mathcal{H}[f]}{p} \dfracp{f}{q}
  - \dfracp{\mathcal{H}[f]}{q} \dfracp{f}{p} = 0,
\end{equation}
where the one-particle distribution function $f(q,p,t)$
is defined on the two-dimensional phase space $\mu=(-\pi,\pi]\times\mathbb{R}$.
The one-particle Hamiltonian functional $\mathcal{H}[f]$ is defined by
\begin{equation}
  \mathcal{H}[f](q,p,t) = \dfrac{p^{2}}{2} + \mathcal{V}[f](q,t),
\end{equation}
where the potential functional is
\begin{equation}
  \mathcal{V}[f](q,t) = \int \Phi(q-q') f(q',p',t) dq'dp'.
\end{equation}
Omitting the sine part in $\mathcal{V}[f]$ again from the rotational
symmetry of the system,
and introducing the order parameter functionals defined by
\begin{equation}
  \mathcal{M}_{a}[f](t) = \iint_{\mu} \cos aq~ f(q,p,t) dqdp, \quad (a=1,2)
\end{equation}
the potential functional is rewritten as
\begin{equation}
  \mathcal{V}[f](q,t) = 1 - \sum_{a=1}^{2} (\Lambda_{a}\mathcal{M}_{a}[f](t)+h_{a})\cos aq.
\end{equation}

\subsection{Three phases in unforced equilibrium state}
\label{sec:unforced-eq} 

The canonical thermal equilibrium states
with zero external fields, $h_{1}=h_{2}=0$, are written as
\begin{equation}
  \label{eq:f0eq}
  \feqz(q,p)
  = \dfrac{G(\Heqz,\beta)}{\iint_{\mu} G(\Heqz,\beta) dqdp},
  \quad
  G(E,\beta) = e^{-\beta E},
\end{equation}
where $\beta$ is the inverse temperature,
\begin{equation}
  \Heqz = \mathcal{H}[\feqz] = \dfrac{p^{2}}{2} + \Veqz(q),
\end{equation}
and
\begin{equation}
  \Veqz(q) = \mathcal{V}[\feqz]
  = 1 - \Lambda_{1}m_{10}\cos q - \Lambda_{2}m_{20}\cos 2q.
\end{equation}
The values of $m_{10}$ and $m_{20}$ are determined by solving the simultaneous
self-consistent equations
\begin{equation}
  \label{eq:self-consistent}
  m_{10} = \mathcal{M}_{1}[\feqz], \quad m_{20} = \mathcal{M}_{2}[\feqz].
\end{equation}
Note that the right-hand-sides depend on $m_{10}$ and $m_{20}$
through $\Heqz$.
We may set $m_{10}$ and $m_{20}$ as non-negative
from the rotational symmetry of the system.

\begin{figure}
  \centering
  \includegraphics[width=8.5cm]{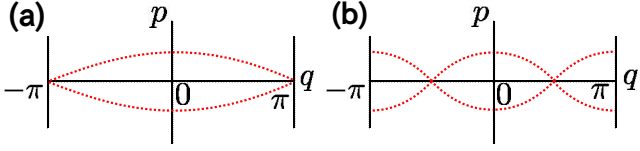}
  \caption{Schematic pictures of phase space.
    The red dotted lines represent separatrices.
    (a) Ferro phase with $m_{10}\gg m_{20}$.
    (b) Nematic phase.}
  \label{fig:phasespace}
\end{figure}

The three phases of the GHMF model are characterized as
\begin{equation}
  \begin{split}
    \text{Para:} & \quad m_{10}=0,~ m_{20}=0, \\
    \text{Nematic:} & \quad m_{10}=0,~ m_{20}>0, \\
    \text{Ferro:} & \quad m_{10}>0,~ m_{20}>0. \\
  \end{split}
\end{equation}
On the two-dimensional phase space $\mu$,
the separatrix is an iso-$\Heqz$ contour,
and forms the skeleton of the phase space.
The Para phase has no separatrix,
since the iso-$\Heqz$ contours in the Para phase
coincide with the iso-$p$ contours.
On the other hand, the Ferro and the Nematic phases have
separatrices as schematically shown in Fig.\ref{fig:phasespace}.

\section{Response in statistical mechanics}
\label{sec:statistical-mechanics}

Before going to the linear response in the Vlasov dynamics,
we revisit the linear response in statistical mechanics for comparison.
The Vlasov dynamics corresponds to the microcanonical ensemble,
but the microcanonical one gives the equivalent phase diagram
with the canonical one except for the parameter region
where the first-order phase transition exists \cite{pikovsky-14}.
We are interested in the susceptibility around the second-order phase transition,
and hence we discuss on the response in the canonical ensemble
for simplicity.

The susceptibility is defined in
Sec. \ref{sec:statistical-mechanics-susceptibility}
by applying constant external fields, $h_{1}$ and $h_{2}$.
The critical lines are discussed in Sec. \ref{sec:critical-lines}
based on the divergence of the susceptibility.
The critical exponent matrices are obtained in Sec. \ref{sec:cr-exp-eq}.

\subsection{Susceptibility}
\label{sec:statistical-mechanics-susceptibility}

Let the system be in the canonical thermal equilibrium state $\feqz$
in the time interval $t<0$.
We apply constant external fields $h_{1}$ and $h_{2}$ at the time $t=0$
and wait a long time.
Then the system is expected to relax to the forced canonical
thermal equilibrium state
\begin{equation}
  f^{\rm c}(q,p)
  = \dfrac{G(H^{\rm c},\beta)}{\iint_{\mu} G(H^{\rm c},\beta) dqdp}
\end{equation}
with the forced equilibrium Hamiltonian $H^{\rm c}=\mathcal{H}[f^{\rm c}]$.
The relaxation is not always true in the Vlasov dynamics,
but we consider the state $f^{\rm c}(q,p)$ in this section.
The order parameters in $f^{\rm c}$ are denoted by
\begin{equation}
  m_{a}^{\rm c} = \mathcal{M}_{a}[f^{\rm c}],  \quad (a=1,2).
\end{equation}

Expanding the order parameters as
\begin{equation}
  m_{a}^{\rm c} = m_{a0} + \delta m_{a}^{\rm c}, \quad (a=1,2),
\end{equation}
the Hamiltonian $H^{\rm c}$ is also expanded as
\begin{equation}
  H^{\rm c} = H_{0} + \delta V^{\rm c}
\end{equation}
with the discrepancy of potential
\begin{equation}
  \delta V^{\rm c}
  = - (\Lambda_{1}\delta m_{1}^{\rm c}+h_{1})\cos q
  - (\Lambda_{2}\delta m_{2}^{\rm c}+h_{2})\cos 2q.
\end{equation}
When the external field $\bh=(h_{1},h_{2})^{\rm T}$ is small,
where the superscript T represents the transposition,
the discrepancy $\delta V^{\rm c}$ is also small
and $f^{\rm c}$ is expanded as
\begin{equation}
  \label{eq:feq-expand}
  f^{\rm c} \simeq f_{0} - \beta f_{0}
  \left( \delta V^{\rm c} - \ave{\delta V^{\rm c}}_{0} \right)
  + O(\norm{\bh}^{2}).
\end{equation}
Here the symbol $\ave{X}_{0}$ represents the average of
the observable $X(q,p)$ over $f_{0}$ as
\begin{equation}
  \ave{X}_{0} = \iint_{\mu} X(q,p) f_{0}(q,p) dqdp. 
\end{equation}
Multiplying \eqref{eq:feq-expand} by $\cos aq$ and integrating over $\mu$,
we have the self-consistent equations and their formal solution
\begin{equation}
  \label{eq:deltam-eq}
  \delta\bbm^{\rm c} = (D^{\rm c})^{-1} \beta C^{\rm c}\bh,
\end{equation}
where the matrix $D^{\rm c}$ is defined by
\begin{equation}
  \label{eq:Deq}
  D^{\rm c} = \mathbb{1}_{2} - \beta C^{\rm c} \Lambda,
\end{equation}
the $(a,b)$-element of the matrix $C^{\rm c}$ is defined by
\begin{equation}
  \label{eq:Ceq}
  C_{ab}^{\rm c} = \ave{\cos aq\cos bq}_{0} - \ave{\cos aq}_{0} \ave{\cos bq}_{0},
\end{equation}
and
\begin{equation}
  \delta\bbm^{\rm c} = 
  \begin{pmatrix}
    \delta m_{1}^{\rm c} \\
    \delta m_{2}^{\rm c} \\
  \end{pmatrix},
  \quad
  \mathbb{1}_{2} =
  \begin{pmatrix}
    1 & 0 \\ 
    0 & 1 \\
  \end{pmatrix},
  \quad
  \Lambda = 
  \begin{pmatrix}
    \Lambda_{1} & 0 \\
    0 & \Lambda_{2} \\
  \end{pmatrix}.
\end{equation}

The susceptibility matrix $\chi^{\rm c}=(\chi^{\rm c}_{ab})$ is defined by
\begin{equation}
  \chi^{\rm c}_{ab} = \lim_{\norm{\bh}\to 0} \dfracp{\delta m^{\rm c}_{a}}{h_{b}},
\end{equation}
and the response formula \eqref{eq:deltam-eq} gives
\begin{equation}
  \label{eq:chieq}
  \chi^{\rm c} = (D^{\rm c})^{-1} \beta C^{\rm c}
  = (D^{\rm c})^{-1} ( \mathbb{1}_{2} - D^{\rm c}) \Lambda^{-1}.
\end{equation}

\subsection{Critical lines}
\label{sec:critical-lines}

Extending the number of order parameters as
\begin{equation}
  \label{eq:m30m40}
  m_{a0} = \iint_{\mu} \cos aq ~\feqz(q,p) dqdp \quad (a=3,4),
\end{equation}
the matrix $C^{\rm c}$ is expressed as
\begin{equation}
  C^{\rm c} = 
  \begin{pmatrix}
    \dfrac{1+m_{20}}{2} - m_{10}^{2} & \dfrac{m_{10}+m_{30}}{2} - m_{10}m_{20} \\
    \dfrac{m_{10}+m_{30}}{2} - m_{10}m_{20} & \dfrac{1+m_{40}}{2} - m_{20}^{2} \\
  \end{pmatrix}.
\end{equation}
On the three critical lines, the order parameter $m_{10}$ is always zero,
which induces $m_{30}=0$ by the parity of the mode numbers,
and the matrix $D^{\rm c}$ can be reduced to
\begin{equation}
  D^{\rm c} =
  \begin{pmatrix}
    1 - \beta \Lambda_{1} \dfrac{1+m_{20}}{2} & 0 \\
    0 & 1 - \beta \Lambda_{2} \left( \dfrac{1+m_{40}}{2} - m_{20}^{2} \right) \\
  \end{pmatrix}.
\end{equation}
The critical point has $\det D^{\rm c}=0$,
which determines the critical inverse temperature $\beta$
for fixed $\Lambda_{1}$ and $\Lambda_{2}$,
or the critical parameter $\Lambda_{1}$ ($\Lambda_{2}$)
for fixed $\beta$ and $\Lambda_{2}$ ($\Lambda_{1}$).

The Para-Ferro and the Nematic-Ferro phase transitions are
ruled by the order parameter $m_{10}$,
and the Para-Nematic phase transition by $m_{20}$.
Therefore, the critical lines are obtained as
\begin{equation}
  \label{eq:critical-lines}
  \begin{split}
    \text{Para-Ferro:} \quad
    & 1 - \dfrac{\beta \Lambda_{1}}{2} = 0, \\
    \text{Nematic-Ferro:} \quad
    & 1 - \dfrac{\beta \Lambda_{1}}{2}(1+m_{20}) = 0, \\
    \text{Para-Nematic:} \quad
    & 1 - \dfrac{\beta\Lambda_{2}}{2} = 0,
  \end{split}
\end{equation}
where we used the fact that $m_{20}=m_{40}=0$
on the critical lines of the Para-Ferro and Para-Nematic phase transitions.
The value of $m_{20}$ in the Nematic-Ferro phase transition are determined
for a given set of $\beta, \Lambda_{1}$ and $\Lambda_{2}$
by solving the self-consistent equations \eqref{eq:self-consistent}
with $m_{10}=0$.

Temperature in the canonical ensemble
can be transformed to energy in the microcanonical ensemble.
The energy functional is defined by
\begin{equation}
  \mathcal{E}[f] = \iint_{\mu} \left(
    \dfrac{p^{2}}{2} + \dfrac{\mathcal{V}[f](q,t)}{2} \right) f(q,p,t) dqdp,
\end{equation}
where the potential is divided by $2$ to avoid the double counting
of pair interactions.
The value of $\mathcal{E}[f]$ is conserved in the Vlasov dynamics.
The unforced equilibrium value of energy $\epsilon_{0}=\mathcal{E}[\feqz]$
is related to the temperature $T=1/\beta$ as
\begin{equation}
  \label{eq:energy-T}
  \epsilon_{0} = \dfrac{T}{2} + \dfrac{1-\Lambda_{1}m_{10}^{2}-\Lambda_{2}m_{20}^{2}}{2}.
\end{equation}
The critical temperature and the critical energy
for a given set of $\Lambda_{1}$ and $\Lambda_{2}$ are arranged
in Table \ref{tab:critical-lines}.

\begin{table}
  \centering
  \begin{tabular}{lll}
    \hline
     & Critical temperature & Critical energy \\
    \hline
    Para-Ferro & $T=\Lambda_{1}/2$ & $\epsilon_{0}=(2+\Lambda_{1})/4$ \\
    Para-Nematic & $T=\Lambda_{2}/2$ & $\epsilon_{0}=(2+\Lambda_{2})/4$ \\
    Nematic-Ferro & $T=\Lambda_{1}(1+m_{20})/2$ & $\epsilon_{0}=\epsilon_{\rm NF}$ \\
    \hline
  \end{tabular}
  \caption{Critical temperature and critical energy
    of second-order phase transitions.
    The critical energy of the Nematic-Ferro phase transition
    is $\epsilon_{\rm NF}=(2+\Lambda_{1})/4 + m_{20}(\Lambda_{1}-2\Lambda_{2}m_{20})/4$.}
  \label{tab:critical-lines}
\end{table}

\subsection{Critical exponent matrix $\gamma^{\rm c}$}
\label{sec:cr-exp-eq}

The critical exponent matrix $\gamma^{\rm c}=(\gamma^{\rm c}_{ab})$
is defined by
\begin{equation}
  \chi^{\rm c}_{ab} \propto \tau^{-\gamma^{\rm c}_{ab}},
\end{equation}
where $\tau$ is the parameter distance from the critical point.
Looking back \eqref{eq:chieq},
we find that the divergences of the susceptibility comes
from the inverse matrix $(D^{\rm c})^{-1}$,
and hence we have to compute $\tau$ dependence of the matrix $D^{\rm c}$.

For later convenience,
we decompose the matrix $D^{\rm c}$ into the two parts as
\begin{equation}
  D^{\rm c} = A^{\rm c} + B^{\rm c},
\end{equation}
where
\begin{equation}
  A^{\rm c} = \mathbb{1}_{2} - \dfrac{\beta}{2}
  \begin{pmatrix}
    1 + m_{20} & m_{10} + m_{30} \\
    m_{10} + m_{30} & 1 + m_{40} \\
  \end{pmatrix}
  \Lambda
\end{equation}
and
\begin{equation}
  B^{\rm c} = \beta
  \begin{pmatrix}
    m_{10}^{2} & m_{10}m_{20} \\
    m_{10}m_{20} & m_{20}^{2} \\
  \end{pmatrix}
  \Lambda.
\end{equation}
As shown later, the $A$ part is common to the Vlasov dynamics,
but the $B$ part is modified.
The estimations of $m_{a0}~(a=1,2,3,4)$ are obtained
from the self-consistent equations for $m_{10}$ and $m_{20}$,
\eqref{eq:self-consistent},
and from the definitions of $m_{30}$ and $m_{40}$,
\eqref{eq:m30m40}.
The analyses are given in the Appendix \ref{sec:estimation-ma0},
and the estimated orders are arranged in Table \ref{tab:order-of-m}.

\begin{table}
  \centering
  \begin{tabular}[c]{c|cccc}
    \hline
    & $m_{10}$ & $m_{20}$ & $m_{30}$ & $m_{40}$ \\
    \hline
    PF & 0 & 0 & 0 & 0 \\
    FP & $O(\tau^{1/2})$ & $O(\tau)$ & $O(\tau^{3/2})$ & $O(\tau^{2})$ \\
    \hline
    PN & 0 & 0 & 0 & 0 \\
    NP & 0 & $O(\tau^{1/2})$ & 0 & $O(\tau)$ \\
    \hline
    NF & 0 & $O(1)$ & 0 & $O(1)$ \\
    FN & $O(\tau^{1/2})$ & $O(1)$ & $O(\tau^{1/2})$ & $O(1)$ \\
    \hline
  \end{tabular}
  \caption{$\tau$ dependence of the spontaneous order parameters
    $m_{10}, m_{20}, m_{30}$ and $m_{40}$,
    where $\tau$ is the parameter distance from the critical point.
    PF and FP represent, for instance, the Para side
    and the Ferro side of the Para-Ferro phase transition,
    respectively.}
  \label{tab:order-of-m}
\end{table}

We may assume, around the critical lines,
the left-hand-sides of \eqref{eq:critical-lines}
are of $O(\tau)$ in general.
This assumption and Table \ref{tab:order-of-m}
give estimations of the matrices $D^{\rm c}$'s as
\begin{equation}
  \begin{split}
    & D^{\rm c}({\rm PF}) =
    \begin{pmatrix}
      O(\tau) & 0 \\ 0 & O(1) 
    \end{pmatrix},
    \quad
    D^{\rm c}({\rm FP}) = 
    \begin{pmatrix}
      O(\tau) & O(\tau^{1/2}) \\ O(\tau^{1/2}) & O(1) 
    \end{pmatrix}, \\
    & D^{\rm c}({\rm PN}) =
    \begin{pmatrix}
      O(1) & 0 \\ 0 & O(\tau)
    \end{pmatrix},
    \quad
    D^{\rm c}({\rm NP}) = 
    \begin{pmatrix}
      O(1) & 0 \\ 0 & O(\tau) 
    \end{pmatrix}, \\
    & D^{\rm c}({\rm NF}) =
    \begin{pmatrix}
      O(\tau) & 0 \\ 0 & O(1)
    \end{pmatrix},
    \quad
    D^{\rm c}({\rm FN}) = 
    \begin{pmatrix}
      O(\tau) & O(\tau^{1/2}) \\ O(\tau^{1/2}) & O(1) 
    \end{pmatrix}, \\
  \end{split}
\end{equation}
where NF and FN represent, for instance,
the Nematic side and the Ferro side
of the Nematic-Ferro phase transition, respectively.
We remark that the orders of elements of the matrix $B^{\rm c}$
are equal to or higher than the matrix $A^{\rm c}$,
and the matrix $B^{\rm c}$ is negligible for computing
the critical exponent matrices in thermal equilibrium.

\begin{figure}
  \centering
  \includegraphics[width=8cm]{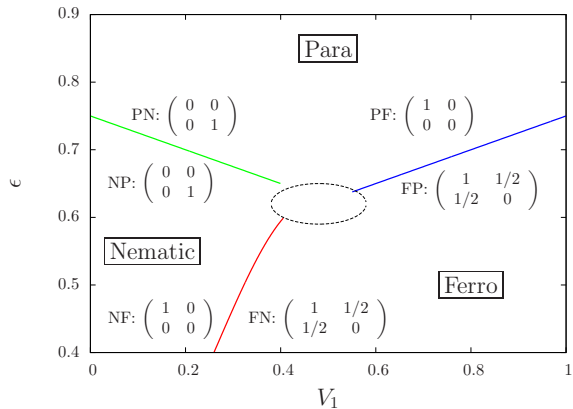}
  \caption{Phase diagram on $(\Lambda_{1},\epsilon)$ plane. $\Lambda_{2}=1-\Lambda_{1}$.
    The region around the black dashed circle
    includes the first-order phase transition \cite{teles-12,pikovsky-14},
    and is out-of-range of the present investigation.
    The blue right, the green left, and red lower lines
    are the critical lines of the Para-Ferro, the Para-Nematic
    and the Nematic-Ferro phase transitions, respectively.
    In each side of the three phase transitions,
    the critical exponent matrix $\gamma^{\rm c}$ is reported.}
  \label{fig:phase-diagram-eq}
\end{figure}

Coming back to the formula \eqref{eq:chieq},
we have the critical exponent matrices as
\begin{equation}
  \label{eq:cr-exp-eq}
  \begin{split}
    & \gamma^{\rm c}({\rm PF}) =
    \begin{pmatrix}
      1 & 0 \\ 0 & 0 \\
    \end{pmatrix},
    \quad
    \gamma^{\rm c}({\rm FP}) =
    \begin{pmatrix}
      1 & 1/2 \\ 1/2 & 0 \\
    \end{pmatrix},\\
    & \gamma^{\rm c}({\rm PN}) =
    \begin{pmatrix}
      0 & 0 \\ 0 & 1 \\
    \end{pmatrix},
    \quad
    \gamma^{\rm c}({\rm NP}) =
    \begin{pmatrix}
      0 & 0 \\ 0 & 1 \\
    \end{pmatrix},\\
    & \gamma^{\rm c}({\rm NF}) =
    \begin{pmatrix}
      1 & 0 \\ 0 & 0 \\
    \end{pmatrix},
    \quad
    \gamma^{\rm c}({\rm FN}) =
    \begin{pmatrix}
      1 & 1/2 \\ 1/2 & 0 \\
    \end{pmatrix}.\\
  \end{split}
\end{equation}
Here we assigned the critical exponent $0$
if no divergence appears.
These critical exponent matrices are reported
in Fig.\ref{fig:phase-diagram-eq}
with the critical lines obtained in Sec.\ref{sec:critical-lines}.

The microcanonical susceptibility does not exceed
the canonical susceptibility due to existence of energy conservation
\cite{mazur-69,suzuki-71},
but the two types of susceptibility
share the critical exponents
as shown in the HMF model \cite{ogawa-patelli-yamaguchi-14}.
See Appendix \ref{sec:micro-cr-exp} for the critical exponents
in the microcanonical ensemble.

\section{Response in Vlasov dynamics}
\label{sec:dynamics}

A nonlinear response theory is recently proposed 
for the Vlasov systems \cite{ogawa-yamaguchi-14}.
Based on it, a simple response formula has been provided \cite{ogawa-yamaguchi-15},
which unifies the nonlinear response theory with the linear response theory
\cite{patelli-gupta-nardini-ruffo-12,ogawa-yamaguchi-12}.
The formula is valid under some conditions,
but they are satisfied for computing the critical exponents
\cite{yamaguchi-ogawa-15}.
We first review the formula 
in Sec.\ref{sec:Vlasov-response-formula},
and the critical exponent matrices are obtained in Sec.\ref{sec:cr-exp-qss}.
We further discuss on negative elements of a susceptibility matrix
in Sec.\ref{sec:negative-susceptibility}.

\subsection{Response formula}
\label{sec:Vlasov-response-formula}

The setting is the same with the case of thermal equilibrium
discussed in Sec.\ref{sec:statistical-mechanics}.
The initial state is the unforced thermal equilibrium state
$\feqz$ \eqref{eq:f0eq}
and then a small external field $\bh$ is applied at the time $t=0$.
Under the Vlasov dynamics, the state, however,
does not go to the forced thermal equilibrium
state $f^{\rm c}$: it is trapped at a nonequilibrium state denoted by $f^{\rm V}$
due to the Casimir invariants of the form $\iint_{\mu} s(f) dqdp$,
where $s$ is an arbitrary differentiable function.
The response formula predicts $f^{\rm V}$ from $\feqz$.

The associated one-particle Hamiltonian $H^{\rm V}=\mathcal{H}[f^{\rm V}]$
is integrable since $f^{\rm V}$ is stationary and $H^{\rm V}$ has
one degree of freedom accordingly.
The integrability introduces the angle-action variables
$(\theta,J)$ associated with $H^{\rm V}$,
and $H^{\rm V}$ can be written as a function of $J$ only.

Roughly speaking, the response formula is expressed as
\cite{ogawa-yamaguchi-14,ogawa-yamaguchi-15}
\begin{equation}
  \label{eq:fqss}
  f^{\rm V} = \ave{\feqz}_{H^{\rm V}},
\end{equation}
where the bracket means the average over the angle variable as
\begin{equation}
  \ave{X}_{H^{\rm V}} = \dfrac{1}{2\pi} \int_{0}^{2\pi} X(q(\theta,J),p(\theta,J)) d\theta.
\end{equation}
In other words, $f^{\rm V}$ is obtained by taking time average
of the initial state $\feqz$
under the Hamiltonian flow associated with $H^{\rm V}$.

For obtaining the response,
as done in Sec.\ref{sec:statistical-mechanics-susceptibility},
we expand the right-hand-side of \eqref{eq:fqss}
with the expansion
\begin{equation}
  \label{eq:HV-Heqz-dVV}
  H^{\rm V} = \Heqz + \delta V^{\rm V}
\end{equation}
where
\begin{equation}
  \delta V^{\rm V}
  = -(\Lambda_{1}\delta m_{1}^{\rm V}+h_{1}) \cos q
  -(\Lambda_{2}\delta m_{2}^{\rm V}+h_{2}) \cos 2q.
\end{equation}
The key idea for expanding the right-hand-side of 
the formula \eqref{eq:fqss} is to use the equality
\begin{equation}
  \ave{\varphi(H^{\rm V})}_{\rm H^{\rm V}} = \varphi(H^{\rm V})
  \quad
  \text{for any } \varphi
\end{equation}
which holds from the definition of the partial average $\ave{\cdot}_{H^{\rm V}}$.
Substituting $\Heqz=H^{\rm V}-\delta V^{\rm V}$ 
into the explicit expression
\begin{equation}
  \label{eq:GH0HV}
  \ave{\feqz}_{H^{\rm V}}
  = \dfrac{\ave{ G(\Heqz,\beta) }_{H^{\rm V}}}{\iint_{\mu} G(\Heqz,\beta) dqdp} 
\end{equation}
and expanding the right-hand-side with respect to the small
$\delta V^{\rm V}$, we have
\begin{equation}
  \label{eq:fqss-expand}
  f^{\rm V} \simeq \feqz
  - \beta f_{0} \left( \delta V^{\rm V} - \ave{\delta V^{\rm V}}_{\Heqz} \right).
\end{equation}
In the way we performed the expansion again
by using $H^{\rm V}=\Heqz+\delta V^{\rm V}$.
The bracket $\ave{\cdot}_{\Heqz}$ is the average
over the angle variable associated with $\Heqz$.
We omitted a higher order contribution
coming from the replacement of
$\ave{\cdot}_{H^{\rm V}}$ with $\ave{\cdot}_{\Heqz}$.

Multiplying \eqref{eq:fqss-expand} by $\cos aq$ 
and integrating over $\mu$,
we have a similar formula for susceptibility
with thermal equilibrium \eqref{eq:chieq} as
\begin{equation}
  \label{eq:chiqss}
  \chi^{\rm V} = (D^{\rm V})^{-1} \beta C^{\rm V}
  = (D^{\rm V})^{-1} ( \mathbb{1}_{2}-D^{\rm V}) \Lambda^{-1}
\end{equation}
but with the different matrix $D^{\rm V}$
\begin{equation}
  D^{\rm V} = \mathbb{1}_{2} - \beta C^{\rm V} \Lambda
\end{equation}
where the $(a,b)$-element of the matrix $C^{\rm V}$ is
\begin{equation}
  \label{eq:Cqss}
  C_{ab}^{\rm V} =  \ave{\cos aq\cos bq}_{0}
  - \ave{\ave{\cos aq}_{\Heqz} \ave{\cos bq}_{\Heqz}}_{0}.
\end{equation}
Here we used the equality
\begin{equation}
  \label{eq:psi1psi2}
  \ave{\psi_{1} \ave{\psi_{2}}_{\Heqz}}_{0}
  = \ave{\ave{\psi_{1}}_{\Heqz}\ave{\psi_{2}}_{\Heqz}}_{0}.
\end{equation}

The matrix $D^{\rm V}$ is decomposed into the two parts as
\begin{equation}
  D^{\rm V} = A^{\rm c} + B^{\rm V}
\end{equation}
where the $(a,b)$-element of the matrix $B^{\rm V}$ is
\begin{equation}
  \begin{split}
  \label{eq:BV-definition}
  B_{ab}^{\rm V}
  & = \beta \ave{\ave{\cos aq}_{\Heqz} \ave{\cos bq}_{\Heqz} }_{0} \Lambda_{b}. \\
  \end{split}
\end{equation}
See the Appendix \ref{sec:integral-BV} for a definite integral formula
of each element in a reduced case.
The matrix $B^{\rm V}$ results to the matrix $B^{\rm c}$
if we replace the partial average over the angle variable,
$\ave{\cos bq}_{\Heqz}$,
with the average over $\feqz$, $\ave{\cos bq}_{0}$.
However, existence of the partial average
modifies the critical exponents.

\subsection{Critical exponent matrix $\gamma^{\rm V}$}
\label{sec:cr-exp-qss}

According to the Appendix \ref{sec:estimation-Dqss},
the matrices $B^{\rm V}$'s are estimated as
\begin{equation}
  \label{eq:BV}
  \begin{split}
    & B^{\rm V}({\rm PF}) =
    \begin{pmatrix}
      0 & 0 \\ 0 & 0 \\
    \end{pmatrix},
    \quad
    B^{\rm V}({\rm FP}) =
    \begin{pmatrix}
      O(\tau^{1/4}) & O(\tau^{1/4}) \\ O(\tau^{1/4}) & O(\tau^{1/4}) \\
    \end{pmatrix},\\
    & B^{\rm V}({\rm PN}) =
    \begin{pmatrix}
      0 & 0 \\ 0 & 0 \\
    \end{pmatrix},
    \quad
    B^{\rm V}({\rm NP}) =
    \begin{pmatrix}
      O(\tau^{1/4}) & 0 \\ 0 & O(\tau^{1/4}) \\
    \end{pmatrix},\\
    & B^{\rm V}({\rm NF}) =
    \begin{pmatrix}
      O(1) & 0 \\ 0 & O(1) \\
    \end{pmatrix},
    \quad
    B^{\rm V}({\rm FN}) =
    \begin{pmatrix}
      O(1) & O(\tau^{c_{1}}) \\ O(\tau^{c_{2}}) & O(1) \\
    \end{pmatrix}.\\
  \end{split}
\end{equation}
The constants $c_{1}$ and $c_{2}$ in $B^{\rm V}({\rm FN})$ are positive
and we skip to compute their precise values
since they do not contribute to the critical exponents as shown later.

\begin{figure}
  \centering
  \includegraphics[width=8cm]{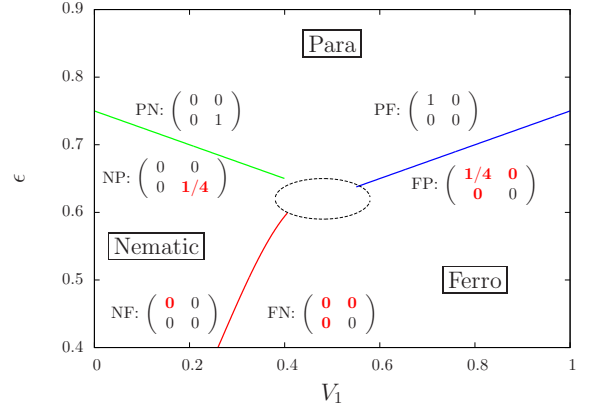}
  \caption{The same with Fig.\ref{fig:phase-diagram-eq}
    but the critical exponent matrices are for the Vlasov dynamics,
    $\gamma^{\rm V}$.
    The red bold elements are different values from the corresponding
    equilibrium values.}
  \label{fig:phase-diagram-dyn}
\end{figure}

As contrasted with thermal equilibrium case,
the matrix $B^{\rm V}$ can partially dominate the matrix $D^{\rm V}$.
This domination modifies the estimations of $D^{\rm V}$'s from $D^{\rm c}$'s as
\begin{equation}
  \label{eq:DV}
  \begin{split}
    & D^{\rm V}({\rm PF}) =
    \begin{pmatrix}
      O(\tau) & 0 \\ 0 & O(1) 
    \end{pmatrix},
    \quad
    D^{\rm V}({\rm FP}) = 
    \begin{pmatrix}
      O(\tau^{1/4}) & O(\tau^{1/4}) \\ O(\tau^{1/4}) & O(1)
    \end{pmatrix}, \\
    & D^{\rm V}({\rm PN}) =
    \begin{pmatrix}
      O(1) & 0 \\ 0 & O(\tau)
    \end{pmatrix},
    \quad
    D^{\rm V}({\rm NP}) = 
    \begin{pmatrix}
      O(1) & 0 \\ 0 & O(\tau^{1/4}) 
    \end{pmatrix}, \\
    & D^{\rm V}({\rm NF}) =
    \begin{pmatrix}
      O(1) & 0 \\ 0 & O(1)
    \end{pmatrix},
    \quad
    D^{\rm V}({\rm FN}) = 
    \begin{pmatrix}
      O(1) & O(\tau^{\bar{c}_{1}}) \\ O(\tau^{\bar{c}_{2}}) & O(1) 
    \end{pmatrix}, \\
  \end{split}
\end{equation}
where $\bar{c}_{j}=\min\{1/2,c_{j}\}~(j=1,2)$.
Recalling the susceptibility formula \eqref{eq:chiqss},
we have the critical exponent matrices $\gamma^{\rm V}$ as
\begin{equation}
  \label{eq:cr-exp-qss}
  \begin{split}
    & \gamma^{\rm V}({\rm PF}) =
    \begin{pmatrix}
      1 & 0 \\ 0 & 0 \\
    \end{pmatrix},
    \quad
    \gamma^{\rm V}({\rm FP}) =
    \begin{pmatrix}
      1/4 & 0 \\ 0 & 0 \\
    \end{pmatrix},\\
    & \gamma^{\rm V}({\rm PN}) =
    \begin{pmatrix}
      0 & 0 \\ 0 & 1 \\
    \end{pmatrix},
    \quad
    \gamma^{\rm V}({\rm NP}) =
    \begin{pmatrix}
      0 & 0 \\ 0 & 1/4 \\
    \end{pmatrix},\\
    & \gamma^{\rm V}({\rm NF}) =
    \begin{pmatrix}
      0 & 0 \\ 0 & 0 \\
    \end{pmatrix},
    \quad
    \gamma^{\rm V}({\rm FN}) =
    \begin{pmatrix}
      0 & 0 \\ 0 & 0 \\
    \end{pmatrix},\\
  \end{split}
\end{equation}
where we assigned the critical exponents $0$
when no divergences appear
even if $\chi_{ab}$'s go to zeros in the limit $\tau\to 0$.
The obtained critical exponent matrices are shown
in Fig.\ref{fig:phase-diagram-dyn}
with stressing the different values from the thermal equilibrium case.

We remark that existence of invariants suppress the response
\cite{mazur-69,suzuki-71},
and hence $\chi^{\rm V}\leq\chi^{\rm c}$.
This fact implies that no critical lines exist on the parameter plane 
except for the ones obtained in canonical statistical mechanics.
Consequently, there are no shifts of the critical lines
and the new zero critical exponents correctly
capture the dynamical obstacle to divergences of susceptibility
at the critical point.

\subsection{Negative susceptibility in  the Para-Ferro phase transition}
\label{sec:negative-susceptibility}
We note that the susceptibility matrix $\chi$ is proportional to
$D^{-1}(\mathbb{1}_{2}-D)=D^{-1}-\mathbb{1}_{2}$, which is written as
\begin{equation}
  D^{-1}-\mathbb{1}_{2} = \dfrac{1}{\det D}
  \begin{pmatrix}
    D_{22} - \det D & - D_{12} \\
    -D_{21} & D_{11} - \det D \\
  \end{pmatrix}.
\end{equation}
We consider if the signs of the off-diagonal elements can change
around the critical point.
The sign of $\det D$ must be fixed in one side of a phase transition
around the critical line, since $\det D=0$ appears only on the critical lines.
Thus, we concentrate on the off-diagonal elements of the matrix $D$
by focusing on the Ferro side of the Para-Ferro phase transition.

The off-diagonal elements of $D^{\rm c}({\rm FP})$ are proportional to
\begin{equation}
  - \dfrac{m_{10}+m_{30}}{2} + m_{10}m_{20},
\end{equation}
and are dominated by the negative first term.
Thus, no change of sign is possible around the critical point
in thermal equilibrium.
Indeed, $\chi_{21}^{\rm c}({\rm FP})\to +\infty$ in the limit $\tau\to 0$
reflecting the positive critical exponent $\gamma_{12}^{\rm c}({\rm FP})=1/2$.

However, in the Vlasov dynamics, 
the off-diagonal elements of $D^{\rm V}({\rm FP})$, which are proportional to
\begin{equation}
  - \dfrac{m_{10}+m_{30}}{2} + \ave{ \ave{\cos q}_{\Heqz} \ave{\cos 2q}_{\Heqz}}_{0},
\end{equation}
may change the signs around the critical point.
The second term is of $O(\tau^{1/4})$
[see $B^{\rm V}({\rm FP})$ in \eqref{eq:BV}],
and is positive by considering iso-$\Heqz$ contour and $\feqz$,
while the first term is negative and is of $O(\tau^{1/2})$.
Thus, the second term can dominate close to the critical point
and change the signs of susceptibility.
We will numerically demonstrate the negative susceptibility,
i.e. $\chi_{21}^{\rm V}({\rm FP})<0$,
in Sec.\ref{eq:Num-negative-susceptibility}.

\section{Numerical tests}
\label{sec:numerics}

The critical exponents $1/4$ of $\gamma^{\rm V}_{11}({\rm FP})$
and $\gamma^{\rm V}_{22}({\rm NP})$
are direct extensions of the HMF model,
whose Ferro phase also has the same critical exponent
\cite{ogawa-patelli-yamaguchi-14}.
$\gamma^{\rm V}_{12}({\rm FN})=\gamma^{\rm V}_{21}({\rm FN})=0$
may associate with $\gamma^{\rm V}_{11}({\rm FN})=0$.
Therefore, interesting exponents are
$\gamma^{\rm V}_{11}({\rm NF})=\gamma^{\rm V}_{11}({\rm FN})=0$
and $\gamma^{\rm V}_{12}({\rm FP})=\gamma^{\rm V}_{21}({\rm FP})=0$.
We confirm 
$\gamma^{\rm V}_{11}({\rm NF})=\gamma^{\rm V}_{11}({\rm FN})=0$
in Sec.\ref{eq:NumNemaFerro}
and $\gamma^{\rm V}_{21}({\rm FP})=0$ in Sec.\ref{eq:NumFerroPara}
by direct numerical simulations of the Vlasov equation
\eqref{eq:Vlasov}.
The negative susceptibility discussed in Sec.\ref{sec:negative-susceptibility}
is also examined in Sec.\ref{eq:Num-negative-susceptibility}.

We perform the semi-Lagrangian simulations \cite{debuyl-10}
with the fixed time step $\Delta t=0.05$.
The phase space $\mu$ is truncated into $(-\pi,\pi]\times [-4,4]$,
and is divided into the grid size $G\times G$.
The initial state is the unforced thermal equilibrium state $\feqz(q,p)$,
\eqref{eq:f0eq}, and a small external field $h_{1}$ is applied
with keeping $h_{2}=0$.
We compute temporal evolution of the order parameter values
$m_{1}=\mathcal{M}_{1}[f]$ and $m_{2}=\mathcal{M}_{2}[f]$
both for $h_{1}=0$ and for $h_{1}>0$,
which are denoted by $m_{a}(t,0)$ and $m_{a}(t,h_{1})$ $(a=1,2)$
respectively at the time $t$.
Then, we observe the normalized discrepancy between the two
to exclude numerical errors for $h_{1}=0$.
That is, we observe the quantities
\begin{equation}
  \chi_{a1}(t) = \dfrac{m_{a}(t,h_{1})-m_{a}(t,0)}{h_{1}},
  \quad (a=1,2).
\end{equation}

The family of initial states is characterized
by the inverse temperature $\beta=1/T$,
but $\beta$ is just a parameter
and is interpreted as energy by the relation \eqref{eq:energy-T}.
All the simulations are performed for the deterministic Vlasov equation
\eqref{eq:Vlasov}, and no thermal noise is introduced.

\subsection{$\gamma^{\rm V}_{11}({\rm NF})=\gamma^{\rm V}_{11}({\rm FN})=0$}
\label{eq:NumNemaFerro}

Following the previous work \cite{teles-12},
we fix energy as $\epsilon=0.55$
to avoid the first-order phase transition region, and vary $\Lambda_{1}$.
The parameter $\tau$ is, therefore, $\tau=|\Lambda_{1}-\Lambda_{1{\rm c}}|$,
where the critical value $V_{1{\rm c}}$ and the value of $m_{20}$
at the critical point are computed as
\begin{equation}
  \label{eq:critical-pt}
  V_{1{\rm c}} \simeq 0.358989, 
  \quad
  m_{20} \simeq 0518977
\end{equation}
for $h_{1}=h_{2}=0$.
We concentrate on the nondivergence of $\chi_{11}$
at the critical point of the Nematic-Ferro phase transition,
which implies $\gamma^{\rm V}_{11}({\rm NF})=\gamma^{\rm V}_{11}({\rm FN})=0$.

The $(1,1)$-element of the matrix $B^{\rm V}({\rm NF})$ is expressed
in the integral form as
\begin{equation}
  \label{eq:BV-11-NF}
  B^{\rm V}_{11}({\rm NF})
  = \dfrac{\sqrt{2\pi\beta \Lambda_{2}m_{20}}}{I_{0}(\beta \Lambda_{2}m_{20})}
  \int_{0}^{1} e^{-\beta \Lambda_{2}m_{20}(2k^{2}-1)} \dfrac{k}{K(k)} dk
\end{equation}
where $I_{0}$ is the zeroth order modified Bessel function,
and $K(k)$ is the complete elliptic integral of the first kind.
$k=0$ and $k=1$ correspond to the minimum energy
and the separatrix energy respectively.
We used the fact that $\ave{\cos q}_{\Heqz}=0$ in the separatrix outside.
See the Appendix \ref{sec:11element-nematic} for the derivation of
\eqref{eq:BV-11-NF}.
Computing the integral numerically, 
we have the values of $D^{\rm V}_{11}({\rm NF})$ and
$\chi_{11}^{\rm V}({\rm NF})$ as
\begin{equation}
  \label{eq:theory}
  D_{11}^{\rm V}({\rm NF}) \simeq 0.936902,
  \quad
  \chi_{11}^{\rm V}({\rm NF}) 
  = \dfrac{1-D_{11}^{\rm V}}{\Lambda_{1} D_{11}^{\rm V}}
  \simeq 0.187603
\end{equation}
at the critical point.

\begin{figure}[h]
  \centering
  \includegraphics[width=8.5cm]{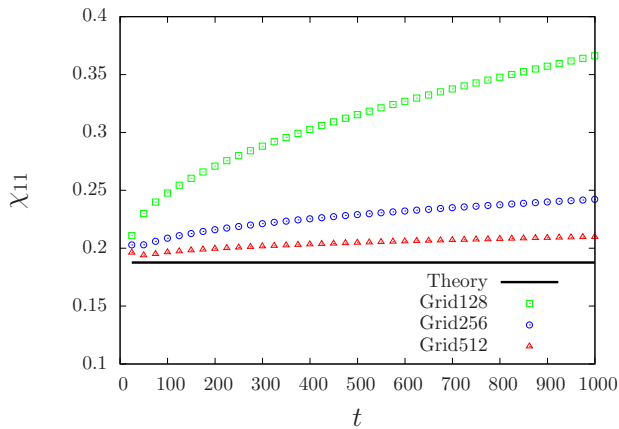}
  \caption{Temporal evolution of $\chi_{11}(t)$
    at the critical point \eqref{eq:critical-pt}
    of the Nematic-Ferro phase transition with energy $\epsilon=0.55$.
    Initial state is $\feqz$, [Eq.\eqref{eq:f0eq}].
    $h_{1}=10^{-4}$ and $h_{2}=0$.
    The grid sizes are $G=128$ (green squares),
    $256$ (blue circles) and $512$ (red triangles).
    The black horizontal solid line is the theoretical
    level of $\chi^{\rm V}_{11}({\rm NF})$, \eqref{eq:theory}.}
  \label{fig:NemaFerroCtPt}
\end{figure}

The grid size dependence of $\chi_{11}(t)$
is reported in Fig.\ref{fig:NemaFerroCtPt},
and the numerical results approach to the theoretical level
as the grid gets finer.
We also computed $h_{1}$ dependence of $\chi_{11}(t)$
with the grid size $G=512$,
and the three numerical curves for $h_{1}=10^{-4},10^{-5}$ and $10^{-6}$ 
almost collapse within the symbol size of Fig.\ref{fig:NemaFerroCtPt}
(not shown).
We, therefore, conclude that the nondivergence of susceptibility
and the finite theoretical level \eqref{eq:theory}
are successfully confirmed at the critical point.

\subsection{$\gamma^{\rm V}_{21}({\rm FP})=0$}
\label{eq:NumFerroPara}

\begin{figure}[h]
  \centering
  \includegraphics[width=8.5cm]{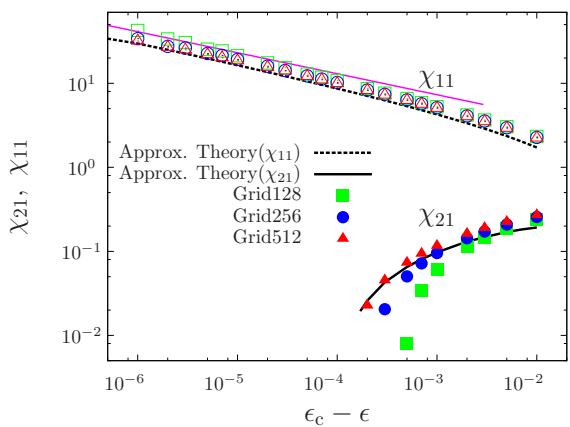}
  \caption{Time averaged $\chi_{11}$ (open symbols)
    and $\chi_{21}$ (filled symbols)
    in the Ferro side of the Para-Ferro phase transitions.
    $h_{1}=10^{-5}$ and $h_{2}=0$.
    The averages are taken in the time window $[200,1000]$.
    The grid size is $G=128$ (green squares), $256$ (blue circles)
    and $512$ (red triangles).
    The black dashed and the black solid lines are susceptibilities
    from an approximate theory for $\chi_{11}$ and $\chi_{21}$ respectively.
    The magenta straight line is a guide of eyes for the slope $-1/4$.}
  \label{fig:FerroPara}
\end{figure}

To avoid the first-order phase transition region again,
we set $\Lambda_{1}=0.8$ and $\Lambda_{2}=0.2$,
which gives the critical energy $\epsilon_{\rm c}=0.7$,
and vary $\epsilon$ below the critical point $\epsilon_{\rm c}$.
Thus, the parameter $\tau$ is $\tau=\epsilon_{\rm c}-\epsilon$,
since we are in the Ferro, low energy phase.
We compute the time averages of $\chi_{11}(t)$ and $\chi_{21}(t)$
in the time window $t\in [200,1000]$.
The averaged susceptibilities are reported in Fig.\ref{fig:FerroPara}
as functions of $\epsilon_{\rm c}-\epsilon$
for the three Grid sizes, $G=128,~256$ and $512$.
The numerical results are in good agreements with an approximate theory,
in which $m_{a0}~(a=2,3,4)$ are neglected
(see the Appendix \ref{sec:integral-BV} for the integral form
of each element of the matrix $B^{\rm c}({\rm FP})$
in this approximated case).
The critical exponent $\gamma^{\rm V}_{11}({\rm FP})=1/4$
is successfully reproduced as in the HMF model
\cite{ogawa-patelli-yamaguchi-14},
and no divergence of $\chi_{21}$ to $+\infty$ is also confirmed.

\subsection{$\chi^{\rm V}_{21}({\rm FP})<0$ close to the critical point}
\label{eq:Num-negative-susceptibility}

\begin{figure}[h]
  \centering
  \includegraphics[width=8.5cm]{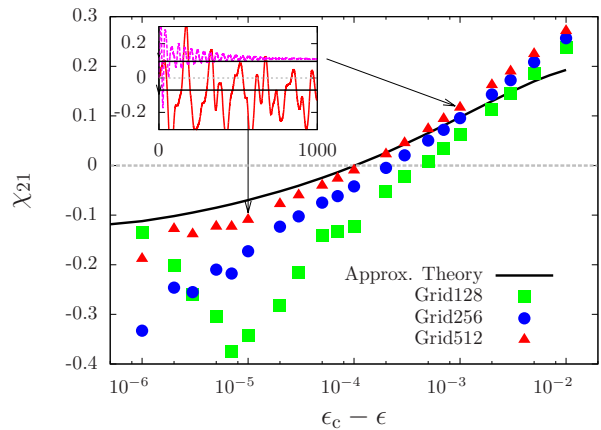}
  \caption{The same with Fig.\ref{fig:FerroPara},
    but we omitted $\chi_{11}$
    and the vertical axis is in the linear scale to observe the negative part.
    (Inset) Temporal evolution of $\chi_{21}(t)$ in the interval $t\in [0,1000]$. The upper magenta and the lower red lines are for $\epsilon_{\rm c}-\epsilon=10^{-3}$ and $10^{-5}$, respectively, with the approximate theoretical levels
    (black solid lines).}
  \label{fig:NegaSusFerroPara}
\end{figure}

The susceptibility $\chi_{21}$ in Fig.\ref{fig:FerroPara} is hidden
close to the critical point,
since $\chi_{21}$ becomes negative.
The negative susceptibility is confirmed
as shown in Fig.\ref{fig:NegaSusFerroPara}
by taking the linear scale for the vertical axis.
This observation in the Vlasov dynamics
gives a sharp contrast with in thermal equilibrium,
in which the susceptibility $\chi_{21}^{\rm c}$
is positive and diverges at the critical point.

\section{Summary and discussions}
\label{sec:summary}

We considered responses to the external fields in the GHMF model.
This model has the two order parameters,
which characterize the Para, the Ferro and the Nematic phases.
In each of thermal equilibrium and of the Vlasov dynamics,
we derived $6$ critical exponent matrices
corresponding to the two sides of the three phase transitions,
where each critical exponent matrix is of $2\times 2$
associated with the two order parameters and their conjugate external fields.

As in the HMF model, the Para phase in the Vlasov dynamics
has the same critical exponent matrices with thermal equilibrium.
This agreement comes from the fact that
both $\ave{\cos aq}_{0}$ and $\ave{\cos aq}_{\Heqz}$
vanish in the Para phase and no dynamical effects appear
in the matrix $D^{\rm V}$.
In the Ferro side of the Para-Ferro phase transition,
and the Nematic side of the Para-Nematic phase transition,
we obtained the suppressed critical exponent $\gamma=1/4$
as the straightforward extension from the HMF model,
where $\gamma=1$ in statistical mechanics.
However, in the Ferro and the Nematic phases,
all the other exponents are zeros,
and no divergences of susceptibility appear at the critical points.
The vanishing critical exponents in the Vlasov dynamics
are stronger suppression than the reduced value of
the previously mentioned $\gamma=1/4$.
These theoretical predictions of no divergences are successfully
confirmed by direct numerical simulations of the Vlasov dynamics.

We found two types of nondivergences of susceptibilities:
one appears in $\chi_{11}^{\rm V}({\rm NF})$ and $\chi_{11}^{\rm V}({\rm FN})$,
and the other in the off-diagonal elements of $\chi^{\rm V}({\rm FP})$.
The former type might be understood by the potential barrier
formed spontaneously by $m_{20}$.
Around the critical point, the potential is $V_{0}\simeq -\Lambda_{2}m_{20}\cos 2q$,
and has the two wells centered at $q=0$ (well-1) and $q=\pi$ (well-2).
Applying the external field to the direction of $q=0$,
particles in the well-2 tend to move to the well-1,
but the potential barrier may prevent them from moving,
since each particle must conserve energy in the Vlasov dynamics.
On the other hand, in the latter type,
the nondivergences comes from the domination of $O(\sqrt{m_{10}})$
in the off-diagonal elements of the matrix $D^{\rm V}$,
but the thermal equilibrium case also have the leading $O(m_{10})$ terms
in the off-diagonal elements of the matrix $D^{\rm c}$.
Thus, the mechanism might not be straightforward comparing with
the former type.

We remark that a non-divergent susceptibility is reported
in the HMF model with a family of spatially homogeneous
but asymmetric momentum distributions \cite{yamaguchi-15}
at the point of stability change.
The thermal equilibrium states, discussed in the present article,
are symmetric and accept non-homogeneous distributions,
and hence the reported non-divergences might have
a larger impact than the asymmetric case.

We also revealed that negative elements appear
in the susceptibility matrix for the Ferro side
of the Para-Ferro phase transition.
The negative susceptibility has been reported in the HMF model
\cite{chavanis-11,deninno-fanelli-12,debuyl-fanelli-ruffo-12}
and in the $\phi^{4}$ model
\cite{campa-ruffo-touchette-07,campa-dauxois-ruffo-09},
but they appear under
the energy conservation between with and without the external field
\cite{chavanis-11}
(see also Appendix \ref{sec:micro-cr-exp}),
the fixed value of magnetization
\cite{campa-ruffo-touchette-07,campa-dauxois-ruffo-09},
or the nonstationary initial states
\cite{deninno-fanelli-12,debuyl-fanelli-ruffo-12}.
The negative susceptibility reported in this article
is observed for the initial thermal equilibrium states
by applying an external field
without any additional constraints, and therefore,
it might be rather easy to compare with experiments.

Finally, it might be worth noting that the nonclassical critical exponents
of the HMF model are also observed in a model of coupled oscillators
by setting the so-called natural frequencies deterministically
\cite{hong-chate-tang-park-15}.
In the model, the oscillators are confined on the unit circle
and the interaction is realized only through the first Fourier mode
as the HMF model.
Thus, it might be interesting to consider a similar extension
in the coupled oscillator system.

\acknowledgements
Y.Y.Y. thanks S.Ogawa for valuable discussions.
He acknowledges the supports of
JSPS KAKENHI Grant Numbers 23560069 and 16K05472.

\appendix

\section{Estimations of $m_{a0}$}
\label{sec:estimation-ma0}
Around the critical point,
we estimate the spontaneous order parameters $m_{a0}$'s,
which are written as
\begin{equation}
  \label{eq:appendix-ma0}
  m_{a0} = \dfrac{\int_{-\pi}^{\pi} \exp[ \beta(\Lambda_{1}m_{10}\cos q + \Lambda_{2}m_{20}\cos 2q) ] \cos aq dq}
  {\int_{-\pi}^{\pi} \exp[ \beta(\Lambda_{1}m_{10}\cos q + \Lambda_{2}m_{20}\cos 2q) ] dq},
\end{equation}
where the denominator is of $O(1)$.
The key idea of this section is to use the orthogonality of
$\{\cos aq\}_{a\in\mathbb{Z}}$, which gives
\begin{equation}
  \label{eq:estimation-m30m40}
  \begin{split}
    & m_{30} = O(m_{10}^{3}) + O(m_{10}m_{20}), \\
    & m_{40} = O(m_{10}^{4}) + O(m_{10}^{2}m_{20}) + O(m_{20}^{2}). \\
  \end{split}
\end{equation}
It is, therefore, enough to estimate $m_{10}$ and $m_{20}$.
We first consider the Para-Ferro and the Para-Nematic phase transitions,
around which $m_{10}$ and $m_{20}$ are small enough,
and then go to the Nematic-Ferro transition.

\subsection{Para-Ferro and Para-Nematic transitions}
All the order parameters are zeros in the Para phase,
and we focus on estimating the order parameters
in the Ferro and the Nematic sides.

The normalization factor, the numerator of the right-hand-side
of \eqref{eq:appendix-ma0}, is expanded as
\begin{equation}
  \begin{split}
    & \int_{-\pi}^{\pi} \exp[ \beta (\Lambda_{1}m_{10}\cos q + \Lambda_{2}m_{20}\cos 2q) ] dq  \\
    & = 2\pi + O(m_{10}^{2}) + O(m_{20}^{2}).
  \end{split}
\end{equation}
Thus, the self-consistent equations,
which are $a=1$ and $2$ in \eqref{eq:appendix-ma0},
are reduced to
\begin{equation}
  \label{eq:self-consistent-expand}
  \begin{split}
    & m_{10} = \dfrac{\beta \Lambda_{1}}{2} m_{10} + O(m_{10}m_{20}) + O(m_{10}^{3}) + \cdots, \\
    & m_{20} = \dfrac{\beta \Lambda_{2}}{2} m_{20} + O(m_{10}^{2}) + O(m_{10}^{2}m_{20}) + O(m_{20}^{3}) + \cdots .
  \end{split}
\end{equation}

In the Ferro side of the Para-Ferro phase transition, 
the ordering is $O(m_{20})\leq O(m_{10}^{2})$ \cite{ogawa-yamaguchi-15},
and $m_{10}$ is determined by the leading two terms as
\begin{equation}
  \left( 1 - \dfrac{\beta \Lambda_{1}}{2} \right) m_{10} + O(m_{10}^{3}) = 0.
\end{equation}
As assumed at the head of Sec.\ref{sec:cr-exp-eq},
the coefficient of the first term is of $O(\tau)$.
Thus, we obtain $m_{10}=O(\tau^{1/2})$
and $m_{20}=O(m_{10}^{2})=O(\tau)$ from the second equation
of \eqref{eq:self-consistent-expand}.
Further, the estimations \eqref{eq:estimation-m30m40} give
$m_{30}=O(\tau^{3/2})$ and $m_{40}=O(\tau^{2})$.

In the Nematic side of the Para-Nematic transition $m_{10}$ is always zero,
and $m_{20}$ is determined by the equation
\begin{equation}
  \left( 1 - \dfrac{\beta \Lambda_{2}}{2} \right) m_{20} + O(m_{20}^{3}) = 0.
\end{equation}
As discussed above, we have $m_{20}=O(\tau^{1/2})$.
Further, $m_{30}=0$ and $m_{40}=O(\tau)$ from \eqref{eq:estimation-m30m40}.

\subsection{Nematic-Ferro transition}
In the Nematic-Ferro transition $m_{20}$ is of $O(1)$,
and we need to estimate $m_{10}$.
Smallness of $m_{10}$ reduces the normalization factor as
\begin{equation}
  \begin{split}
    & \int_{-\pi}^{\pi} \exp[\beta (\Lambda_{1}m_{10}\cos q + \Lambda_{2}m_{20}\cos 2q) ] dq \\
    & \simeq \int_{-\pi}^{\pi} \exp(\beta \Lambda_{2}m_{20}\cos 2q) dq + O(m_{10}^{2}).
  \end{split}
\end{equation}
The self-consistent equation for $m_{10}$ is expanded as
\begin{equation}
  m_{10} = \dfrac{\beta \Lambda_{1}m_{10}}{2} \left ( 1 +
    \dfrac{\int \exp(\beta \Lambda_{2}m_{20}\cos 2q) \cos 2q dq}
    {\int \exp(\beta \Lambda_{2}m_{20}\cos 2q) dq}
  \right)
  + O(m_{10}^{3}),
\end{equation}
and the definition of $m_{20}$ gives
\begin{equation}
  \label{eq:appendix-nemaferro-m10}
  \left[ 1 - \dfrac{\beta \Lambda_{1}m_{10}}{2}(1+m_{20}) \right] m_{10}
  + O(m_{10}^{3}) = 0
\end{equation}
where we used the fact
\begin{equation}
  m_{20}
  = \dfrac{\int \exp [\beta \Lambda_{2}m_{20}\cos 2q ] \cos 2q dq}
  {\int \exp [\beta \Lambda_{2}m_{20}\cos 2q) ] dq} + O(m_{10}^{2}).
\end{equation}
Recalling the critical line \eqref{eq:critical-lines},
from \eqref{eq:appendix-nemaferro-m10},
we conclude that $m_{10}=O(\tau^{1/2})$.
From \eqref{eq:estimation-m30m40} we also estimate
$m_{30}=O(\tau^{1/2})$ and $m_{40}=O(1)$.

\section{Critical exponents in microcanonical ensemble}
\label{sec:micro-cr-exp}

In the microcanonical ensemble,
temperature may be modified by applying an external field $\bh$
at the time $t=0$ due to the energy conservation.
Denoting the modified temperature by $T^{\rm mc}$,
we consider the energy conservation relation
between $t=0^{+}$ and $t\to\infty$ as
\begin{equation}
  \label{eq:energy-conservation}
  \begin{split}
    & \dfrac{T}{2}
    + \dfrac{1-\Lambda_{1}m_{10}^{2}-\Lambda_{2}m_{20}^{2}}{2}
    - h_{1}m_{10} - h_{2}m_{20} \\
    & = \dfrac{T^{\rm mc}}{2}
    + \dfrac{1-\Lambda_{1}(m_{1}^{\rm mc})^{2}-\Lambda_{2}(m_{2}^{\rm mc})^{2}}{2}
    - h_{1}m_{1}^{\rm mc} - h_{2}m_{2}^{\rm mc} \\
  \end{split}
\end{equation}
where $m_{1}^{\rm mc}$ and $m_{2}^{\rm mc}$ are the values of
order parameters in the microcanonical ensemble.
Introducing the response
\begin{equation}
  \delta\bbm^{\rm mc} =
  \begin{pmatrix}
    m_{1}^{\rm mc} - m_{10} \\
    m_{2}^{\rm mc} - m_{20} \\
  \end{pmatrix},
\end{equation}
which will be determined later,
the above relation gives the shift of inverse temperature
from $\beta$ to $\beta+\delta\beta$, where 
\begin{equation}
  \label{eq:deltabeta}
  \delta\beta \simeq - 2\beta^{2}~ \bbm_{0}^{\rm T}~ \Lambda~ \delta\bbm^{\rm mc}
\end{equation}
up to the leading order.

Let us introduce the vectors
\begin{equation}
  \bbm_{0} =
  \begin{pmatrix}
    m_{10} \\ m_{20}
  \end{pmatrix},
  \quad
  \bc(q) =
  \begin{pmatrix}
    \cos q \\ \cos 2q
  \end{pmatrix}
\end{equation}
and the discrepancy of potential
\begin{equation}
  \label{eq:deltaVmc}
  \delta V^{\rm mc} = -\bc^{\rm T}\Lambda~ \delta\bbm^{\rm mc} - \bc^{\rm T}\bh.
\end{equation}
The self-consistent equation in the microcanonical ensemble
is obtained by replacing $\beta\delta V^{\rm c}$
with $\beta\delta V^{\rm mc}+H_{0}\delta\beta$ in \eqref{eq:feq-expand},
\begin{equation}
  \label{eq:micro-self-consistent}
  \delta\bbm^{\rm mc}
  = - \ave{ \bc (\beta\delta V^{\rm mc} + H_{0}\delta\beta)}_{0}
  + \ave{\bc}_{0} \ave{\beta\delta V^{\rm mc} + H_{0}\delta\beta}_{0}.
\end{equation}
Noting that $p^{2}/2$ term of $H_{0}$ is canceled between the two terms,
and substituting \eqref{eq:deltabeta}, \eqref{eq:deltaVmc} and
\begin{equation}
  V_{0} = - \bc^{\rm T} \Lambda \bbm_{0}
\end{equation}
into the self-consistent equation \eqref{eq:micro-self-consistent},
we have
\begin{equation}
  \delta\bbm^{\rm mc} = \beta C^{\rm c}
  \left( \mathbb{1}_{2} - 2\beta \Lambda\bbm_{0}\bbm_{0}^{\rm T} \right) \Lambda \delta\bbm^{\rm mc}
  + \beta C^{\rm c} \bh
\end{equation}
where $C^{\rm c}=\ave{\bc\bc^{\rm T}}_{0}$ by the definition.
The response $\delta\bbm^{\rm mc}$ in the microcanonical ensemble is, therefore,
\begin{equation}
  \delta\bbm^{\rm mc} = (D^{\rm mc})^{-1} \beta C^{\rm c}\bh,
\end{equation}
where the matrix $D^{\rm mc}$ is defined by
\begin{equation}
  D^{\rm mc} = \mathbb{1}_{2} - \beta C^{\rm c}
  \left( \mathbb{1}_{2} - 2\beta \Lambda\bbm_{0}\bbm_{0}^{\rm T} \right) \Lambda.
\end{equation}
The matrix $D^{\rm mc}$ is expressed as
\begin{equation}
  D^{\rm mc} = D^{\rm c} + 2\beta^{2} C^{\rm c} \Lambda \bbm_{0}\bbm_{0}^{\rm T}\Lambda,
\end{equation}
and the second term of the right-hand-side does not change
the dominating $\tau$ dependence
of $D^{\rm mc}$ from $D^{\rm c}$.
This concludes that the critical exponents are shared
between the canonical and the microcanonical ensembles.

We give a remark on usage of the energy conservation.
If we require the energy conservation
between $t=0^{-}$ and $t\to\infty$,
the energy conservation relation is modified
from \eqref{eq:energy-conservation} to
\begin{equation}
  \label{eq:energy-conservation-Efix}
  \begin{split}
    & \dfrac{T}{2}
    + \dfrac{1-\Lambda_{1}m_{10}^{2}-\Lambda_{2}m_{20}^{2}}{2} \\
    & = \dfrac{T^{\rm ene}}{2}
    + \dfrac{1-\Lambda_{1}(m_{1}^{\rm ene})^{2}-\Lambda_{2}(m_{2}^{\rm ene})^{2}}{2}
    - h_{1}m_{1}^{\rm ene} - h_{2}m_{2}^{\rm ene} \\
  \end{split}
\end{equation}
where all the superscripts are replaced to represent
the considering situation.
Then, $\delta\beta$ is modified to
\begin{equation}
  \delta\beta^{\rm ene}
  \simeq - 2\beta^{2} \bbm_{0}^{\rm T} \left(
    \Lambda\delta\bbm^{\rm ene} + \bh \right).
\end{equation}
In the previous setting, the last term was not proportional to
$\bbm_{0}^{\rm T}\bh$ but to $\delta\bbm^{\rm T}\bh$
and was omitted since it was of higher order.
With the modified $\delta\beta^{\rm ene}$, we have the linear response
$\delta\bbm^{\rm ene}$ as
\begin{equation}
  \delta\bbm^{\rm ene}
  = (D^{\rm mc})^{-1} \beta C^{\rm c} \left(
    \mathbb{1}_{2}-2\beta\Lambda\bbm_{0}\bbm_{0}^{\rm T} \right) \bh.
\end{equation}
Divergences of the linear response come from $(D^{\rm mc})^{-1}$ again,
and hence the critical exponents are not modified.
On the other hand, the response $\delta\bbm^{\rm ene}$ may be negative
in the off-diagonal elements
due to the factor $\mathbb{1}_{2}-2\beta\Lambda\bbm_{0}\bbm_{0}^{\rm T}$,
and even in the diagonal elements for large $\beta$,
which implies large $\bbm_{0}$.

\section{Integral formula for elements of the matrix $B^{\rm V}$ in a reduced case}
\label{sec:integral-BV}

We give a useful formula of the matrix $B^{\rm V}$
in the case that the one-particle Hamiltonian is written in the form
\begin{equation}
  \label{eq:AppH0}
  \Heqz(q,p) = \dfrac{p^{2}}{2} - M\cos q,
  \quad (M>0).
\end{equation}
This form includes the Nematic phase by replacing $q$ with $2q$
and setting $M=\Lambda_{2}m_{20}$,
and the approximate theory used in Sec.\ref{eq:NumFerroPara},
which is for the Ferro side of the Para-Ferro phase transition,
by setting $m_{20}=0$ and $M=\Lambda_{1}m_{10}$.
We note that this Hamiltonian is symmetric with respect to
both $q$ and $p$, namely $\Heqz(-q,p)=\Heqz(q,-p)=\Heqz(q,p)$.
We will use this symmetry for reducing computations.

\subsection{Angle-action variables and elliptic integrals}

The Hamiltonian system $\Heqz$ \eqref{eq:AppH0} is integrable,
and we can introduce the associated angle-action variables $(\theta,J)$.
They are written in the use of the Legendre elliptic integrals,
defined by
\begin{equation}
  F(\phi,k) = \int_{0}^{\phi}\dfrac{d\varphi}{\sqrt{1-k^{2}\sin^{2}\varphi}}
  = \int_{0}^{\sin\phi} \dfrac{du}{\sqrt{(1-u^{2})(1-k^{2}u^{2})}}
\end{equation}
and
\begin{equation}
  E(\phi,k) = \int_{0}^{\phi} \sqrt{1-k^{2}\sin^{2}\varphi} d\varphi
  = \int_{0}^{\sin\phi} \sqrt{\dfrac{1-k^{2}u^{2}}{1-u^{2}}} du.
\end{equation}
The complete elliptic integrals of the first and the second kinds
are defined respectively as
\begin{equation}
  K(k) = F(\pi/2,k), \quad E(k) = E(\pi/2,k). 
\end{equation}

The Hamiltonian system $\Heqz$ \eqref{eq:AppH0} has two hyperbolic
fixed points, $(q,p)=(-\pi,0)$ and $(\pi,0)$,
and they are connected by the separatrix.
The phase space $\mu$ is divided into outside and inside of the separatrix.
See Fig.\ref{fig:phasespace}(a) for a schematic picture of the phase space.
Based on this knowledge, the angle-action variables $(\theta,J)$
are introduced as \cite{barre-olivetti-yamaguchi-10}
\begin{equation}
  J = \left\{
    \begin{array}{ll}
      \dfrac{4\sqrt{M}}{\pi} k E(1/k) & \text{separatrix outside} \\
      \dfrac{8\sqrt{M}}{\pi} [E(k)-(1-k^{2})K(k)] & \text{separatrix inside} \\
    \end{array}
  \right.
\end{equation}
and
\begin{equation}
  \theta = \left\{
    \begin{array}{ll}
      \pi \dfrac{F(q/2, 1/k)}{K(1/k)} & \text{separatrix outside, upper-half} \\
      \dfrac{\pi}{2} \dfrac{F(Q,k)}{K(k)} & \text{separatrix inside, upper-half} \\
      \dfrac{\pi}{2} \left( 2 - \dfrac{F(Q,k)}{K(k)} \right) & \text{separatrix inside, lower-half} \\
      - \pi \dfrac{F(q/2, 1/k)}{K(1/k)} & \text{separatrix outside, lower-half} \\
    \end{array}
  \right.
\end{equation}
where
\begin{equation}
  \label{eq:appendix-k}
  k = \sqrt{\dfrac{\Heqz(q,p)+M}{2M}}
\end{equation}
and $Q$ is defined by
\begin{equation}
  k\sin Q = \sin(q/2).
\end{equation}
The energy minimum corresponds to $k=0$,
and the separatrix energy to $k=1$.

For later convenience, we introduce the integrals
\begin{equation}
  N_{n}(k) = \int_{0}^{1} \dfrac{u^{n}du}{\sqrt{(1-u^{2})(1-k^{2}u^{2})}}.
\end{equation}
This integrals have the recursive formula
\begin{equation}
  (m+2)k^{2}N_{m+3}(k) - (m+1)(1+k^{2})N_{m+1}(k) + mN_{m-1}(k) = 0,
\end{equation}
and hence
\begin{equation}
  \label{eq:In-integrals}
  \begin{split}
    N_{0}(k) & = K(k), \\
    N_{2}(k) & = \dfrac{K(k)-E(k)}{k^{2}}, \\
    N_{4}(k) & = \dfrac{(2+k^{2})K(k)-2(1+k^{2})E(k)}{3k^{4}}. \\
  \end{split}
\end{equation}

\subsection{Computations of $\ave{\cos bq}_{\Heqz}$}
Let us compute the averages
\begin{equation}
  \ave{\cos^{n}q}_{\Heqz} = \dfrac{1}{2\pi} \int_{-\pi}^{\pi} \cos^{n}q(\theta,J) d\theta.
\end{equation}
Using the elliptic function ${\rm sn}$, which is the inverse function
of $F(\phi,k)$ and is defined by
\begin{equation}
  {\rm sn}(F(\phi,k),k) = \sin\phi, 
\end{equation}
we can write $\cos q$ as \cite{ogawa-yamaguchi-14}
\begin{equation}
  \label{eq:appendix-cosq}
  \cos q = \left\{
    \begin{array}{ll}
      1 - 2{\rm sn}^{2}\left( \dfrac{K(1/k)}{\pi} \theta, \dfrac{1}{k} \right) & \text{separatrix outside,} \\
      1 - 2k^{2}{\rm sn}^{2}\left( \dfrac{2K(k)}{\pi} \theta, k \right) & \text{separatrix inside.} \\
    \end{array}
  \right.
\end{equation}
Changing the variable as
\begin{equation}
  \theta = \left\{
    \begin{array}{ll}
      \dfrac{\pi}{K(1/k)} F(\phi,1/k) & \text{separatrix outside,} \\
      \dfrac{\pi}{2K(k)} F(\phi,k) & \text{separatrix inside,} \\
    \end{array}
  \right.
\end{equation}
and using the symmetry of $\Heqz(q,p)$,
we have the expressions of $\ave{\cos^{n}q}_{\Heqz}$ as
\begin{equation}
  \ave{\cos^{n}q}_{\Heqz}   = \left\{
    \begin{array}{l}
      \dfrac{1}{K(1/k)} \displaystyle{\int_{0}^{1}} \dfrac{(1-2u^{2})^{n}}{\sqrt{(1-u^{2})(1-k^{-2}u^{2})}} du \\
      \text{separatrix outside}, \\
      \dfrac{1}{K(k)} \displaystyle{\int_{0}^{1}} \dfrac{(1-2k^{2}u^{2})^{n}}{\sqrt{(1-u^{2})(1-k^{2}u^{2})}} du \\
      \text{separatrix inside}. \\
    \end{array}
  \right.
\end{equation}
Thus, the integrals $N_{n}(k)$, \eqref{eq:In-integrals}, derive
$\ave{\cos q}_{\Heqz}$ and $\ave{\cos^{2}q}_{\Heqz}$.
The concrete expressions of $\ave{\cos q}_{\Heqz}$ and $\ave{\cos 2q}_{\Heqz}$,
which can be computed from $\ave{\cos^{2}q}_{\Heqz}$, are
\begin{equation}
  \ave{\cos q}_{\Heqz} = \left\{
     \begin{array}{ll}
       2k^{2}\dfrac{E(1/k)}{K(1/k)} - (2k^{2}-1) & \text{separatrix outside}, \\
       2\dfrac{E(k)}{K(k)} - 1 & \text{separatrix inside},
     \end{array}
  \right.
\end{equation}
and
\begin{equation}
  \ave{\cos 2q}_{\Heqz} = \left\{
    \begin{array}{l}
      \dfrac{8k^{2}}{3}(1-2k^{2})\dfrac{E(1/k)}{K(1/k)}
      +1 - \dfrac{16k^{2}}{3}(1-k^{2}) \\
      \text{separatrix outside},\\
      \dfrac{8}{3}(1-2k^{2}) \dfrac{E(k)}{K(k)} - \dfrac{5-8k^{2}}{3} \\
      \text{separatrix inside}.
    \end{array}
  \right.
\end{equation}

\subsection{Computations of $\ave{\ave{\cos aq}_{\Heqz} \ave{\cos bq}_{\Heqz}}_{0}$}

We first show the equality \eqref{eq:psi1psi2}.
Noting that $f_{0}$ and $\ave{\psi_{2}}_{\Heqz}$ depends on $J$ only,
and using $dqdp=d\theta dJ$, we can show 
\begin{equation}
  \begin{split}
    & \ave{ \psi_{1} \ave{\psi_{2}}_{\Heqz} }_{0}
    = \int dJ f_{0} \ave{\psi_{2}}_{\Heqz} \int d\theta \psi_{1} \\
    & = 2\pi \int dJ f_{0} \ave{\psi_{2}}_{\Heqz} \ave{\psi_{1}}_{\Heqz}
    = \iint_{\mu} f_{0} \ave{\psi_{1}}_{\Heqz} \ave{\psi_{2}}_{\Heqz} dqdp \\
    & = \ave{\ave{\psi_{1}}_{\Heqz} \ave{\psi_{2}}_{\Heqz}}_{0}.
  \end{split}
\end{equation}

We then consider the average
\begin{equation}
  \ave{ \ave{\cos aq}_{\Heqz} \ave{\cos bq}_{\Heqz} }_{0}
  = 2\pi \int f_{0} \ave{\cos aq}_{\Heqz} \ave{\cos bq}_{\Heqz} dJ.
\end{equation}
As shown previously,
the average $\ave{\cos aq}_{\Heqz}$ is obtained as a function of $k$,
and accordingly, we change the integral variable from $J$ to $k$
by using the Jacobian
\begin{equation}
  \dfrac{dJ}{dk} = \left\{
    \begin{array}{ll}
      \dfrac{4\sqrt{M}}{\pi} K(1/k) & \text{separatrix outside}, \\ 
      \dfrac{8\sqrt{M}}{\pi} k K(k) & \text{separatrix inside}, \\ 
    \end{array}
  \right.
\end{equation}
where we used the derivatives of $K(k)$ and $E(k)$
\begin{equation}
  \begin{split}
    \dfrac{dK}{dk}(k) & = \dfrac{E(k)-(1-k^{2})K(k)}{k(1-k^{2})}, \\
    \dfrac{dE}{dk}(k) & = \dfrac{E(k)-K(k)}{k}.
  \end{split}
\end{equation}
Denoting the initial stationary state as $\feqz(q,p)=G_{0}(k)$,
and recalling that the separatrix outside has two contributions from the upper
and the lower of the separatrix, we have
\begin{equation}
  \label{eq:avecosaqcosbq-qss}
  \begin{split}
    & \ave{ \ave{\cos aq}_{\Heqz} \ave{\cos bq}_{\Heqz}}_{0} \\
    & = 16 \sqrt{M} \int_{1}^{\infty} G_{0}(k) 
    K(1/k) \ave{\cos aq}_{\Heqz} \ave{\cos bq}_{\Heqz} dk  \\
    & + 16 \sqrt{M} \int_{0}^{1} G_{0}(k) 
    k K(k) \ave{\cos aq}_{\Heqz} \ave{\cos bq}_{\Heqz} dk.
  \end{split}
\end{equation}

\subsection{The $(1,1)$-element in the Nematic phase}
\label{sec:11element-nematic}
We give the $(1,1)$-element in the Nematic phase.
We derive it via replacing $q$ with $2q$ in the obtained results.
We note that the same formula is also derived by starting from the Hamiltonian
\begin{equation}
  H_{0} = \dfrac{p^{2}}{2} - \Lambda_{2}m_{20}\cos 2q.
\end{equation}
The Nematic phase has two hyperbolic fixed points of
$(q,p)=(-\pi/2,0)$ and $(\pi/2,0)$ and the separatrix connects them
by forming two ``eyes'' centered at $(q,p)=(0,0)$ (eye-1)
and $(\pi,0)$ (eye-2).
See Fig.\ref{fig:phasespace}(b).

From symmetry, the average $\ave{\cos q}_{\Heqz}$ is canceled
in separatrix outside. Indeed, the average is modified as
\begin{equation}
  \begin{split}
    & 2\pi \ave{\cos aq}_{\Heqz} = \int_{0}^{2\pi} \cos aq(\theta) d\theta
    = \int_{0}^{2\pi} \cos aq(\theta+\pi) d\theta \\
    & = \int_{0}^{2\pi} \cos a(q(\theta)+\pi) d\theta
    = \cos a\pi \times 2\pi \ave{\cos aq}_{\Heqz}.
\end{split}
\end{equation}
In the eye inside, we have the transform
\begin{equation}
  \cos q
  = \left\{
    \begin{array}{ll}
      \sqrt{\dfrac{1+\cos 2q}{2}} & q\in (-\pi/2,\pi/2) \text{: eye-1} \\
      - \sqrt{\dfrac{1+\cos 2q}{2}} & q\in (-\pi,-\pi/2)\cup(\pi/2,\pi) \text{: eye-2} \\
    \end{array}
  \right.
\end{equation}
and, referring to \eqref{eq:appendix-cosq}, $\cos 2q$ is expressed as
\begin{equation}
  \cos 2q = 1 - 2k^{2} {\rm sn}^{2}\left( \dfrac{2K(k)}{\pi}\theta, k \right).
\end{equation}
Therefore, we totally have
\begin{equation}
  \ave{\cos q}_{\Heqz} = \left\{
    \begin{array}{ll}
      0 & \text{separatrix outside}, \\
      \dfrac{\pi}{2K(k)} & \text{eye-1 inside}. \\
      - \dfrac{\pi}{2K(k)} & \text{eye-2 inside}. \\
    \end{array}
  \right.
\end{equation}

We have two contributions from the eye-1 and the eye-2,
but the factor $2$ is canceled
with the factor $1/2$ from the Jacobian $dJ/dk$. 
Indeed, the action variable defined as
\begin{equation}
  J = \oint p dq
\end{equation}
becomes half since the traveling distance of a periodic orbit
becomes half in the separatrix inside of Fig.\ref{fig:phasespace}(b)
comparing with Fig.\ref{fig:phasespace}(a).
We remark that the action in the separatrix outside
does not change since the traveling distance does not change.

Putting all together with the thermal equilibrium state
\begin{equation}
  \feqz(q,p)
  = \dfrac{e^{-\beta (p^{2}/2-M\cos 2q)}}{\iint_{\mu} e^{-\beta (p^{2}/2-M\cos 2q)} dqdp}
  = \dfrac{e^{-\beta M(2k^{2}-1)}}{\sqrt{2\pi/\beta}~ 2\pi I_{0}(\beta M)},
\end{equation}
we have
\begin{equation}
  \ave{ \ave{\cos q}_{\Heqz} \ave{\cos q}_{\Heqz}}_{0}
  = \dfrac{\sqrt{2\pi\beta M}}{I_{0}(\beta M)}
  \int_{0}^{1} e^{-\beta M(2k^{2}-1)} \dfrac{k}{K(k)} dk.
\end{equation}
This expression results to $B^{\rm V}_{11}({\rm NF})$ \eqref{eq:BV-11-NF}
by setting $M=\Lambda_{2}m_{20}$.

\section{Estimations of $B^{\rm V}$ matrix}
\label{sec:estimation-Dqss}
We give estimations of the matrix $B^{\rm V}$ \eqref{eq:BV-definition}
by using the formula \eqref{eq:avecosaqcosbq-qss},
which implies $B^{\rm V}=O(\sqrt{M})$ by appropriately setting $M$,
since the integral part does not vanish even in the limit $M\to 0$.
Keeping this ordering in mind,
we separately discuss on the Para, the Nematic and the Ferro phases.

\subsection{Para phase}
All the order parameters are zeros in the Para phase,
and the angle variable is nothing but $q$.
Thus, we have $\ave{\cos aq}_{\Heqz}=0$ and hence
\begin{equation}
  B^{\rm V}({\rm PF}) =
  \begin{pmatrix}
    0 & 0 \\
    0 & 0 \\
  \end{pmatrix},
  \quad
  B^{\rm V}({\rm PN}) = 
  \begin{pmatrix}
    0 & 0 \\
    0 & 0 \\
  \end{pmatrix}.
\end{equation}
This is consistent with setting $M=0$
in the formula \eqref{eq:avecosaqcosbq-qss}.

\subsection{Nematic phase}
The parameter $M$ is regarded as $\Lambda_{2}m_{20}$,
and the matrices $B^{\rm V}$ is of $O(m_{20})$.
However, the off-diagonal elements vanish due to cancellation.
The cancellation can be found as follows.
In separatrix outside, we recall $\ave{\cos q}_{\Heqz}=0$
and there is no contribution from the separatrix outside
to the off-diagonal elements.
In separatrix inside, we have contributions from two eyes
(see the Appendix \ref{sec:11element-nematic}).
The contribution from the eye-2 is obtained by shifting
the variable $q$ with $\pi$ in the contribution from the eye-1,
and is multiplied by $\cos a\pi$.
Thus, the total contribution from the two eyes
has the prefactor $1+\cos a\pi\cos b\pi$,
and vanishes for $(a,b)=(1,2)$ and $(2,1)$.
The ordering of $m_{20}$ is $m_{20}=O(\tau^{1/2})$
for the Para-Nematic phase transition,
and $m_{20}=O(1)$ for the Nematic-Ferro phase transition.
These estimations give
\begin{equation}
  B^{\rm V}({\rm NP}) =
  \begin{pmatrix}
    O(\tau^{1/4}) & 0 \\
    0 & O(\tau^{1/4}) \\
  \end{pmatrix},
  \quad
  B^{\rm V}({\rm NF}) = 
  \begin{pmatrix}
    O(1) & 0 \\
    0 & O(1) \\
  \end{pmatrix}.
\end{equation}

\subsection{Ferro phase}

For the Para-Ferro phase transition, we may approximate
the potential as
\begin{equation}
  \Veqz \simeq - \Lambda_{1}m_{10}\cos q,
\end{equation}
and hence the parameter $M$ is regarded as $\Lambda_{1}m_{10}$
and is of $O(\tau^{1/2})$.
There is no reason of cancellation which occurs in the Nematic phase,
and hence we have
\begin{equation}
  B^{\rm V}({\rm FP}) = 
  \begin{pmatrix}
    O(\tau^{1/4}) & O(\tau^{1/4}) \\
    O(\tau^{1/4}) & O(\tau^{1/4}) \\
  \end{pmatrix}.
\end{equation}

For the Nematic-Ferro phase transition,
we may approximate the potential as
\begin{equation}
  \Veqz \simeq - \Lambda_{2}m_{20}\cos 2q,
\end{equation}
and hence the parameter $M$ is regarded as $\Lambda_{2}m_{20}$ and is of $O(1)$.
The approximated potential is the same with one in the Nematic phase,
but the cancellation does not exactly occur by symmetry breaking
due to non-zero $m_{10}$.
The off-diagonal elements may be non-zeros and tend to vanish
as approaching to the critical line.
However, the off-diagonal elements are not important to observe
nondivergence of susceptibility at the critical point of
the Nematic-Ferro phase transition,
and we skip the precise computations.
Consequently, we have
\begin{equation}
  B^{\rm V}({\rm FN}) = 
  \begin{pmatrix}
    O(1) & O(\tau^{c_{1}}) \\
    O(\tau^{c_{2}}) & O(1) \\
  \end{pmatrix}
\end{equation}
with $c_{1},c_{2}>0$.

\end{document}